\documentclass[11pt]{article}
\usepackage[margin=1in]{geometry}
\usepackage{amsmath,amssymb,amsfonts,amsthm}
\usepackage{graphicx}
\usepackage{booktabs}
\usepackage[numbers]{natbib}
\usepackage{hyperref}
\usepackage{enumitem}
\usepackage{microtype}
\usepackage{xcolor}
\usepackage{orcidlink}
\newcommand{\Fourier}{\mathcal{F}}
\newcommand{\InvFourier}{\mathcal{F}^{-1}}
\newcommand{\asm}{\mathcal{P}}
\newcommand{\phasemap}{\Phi}
\newcommand{\measureop}{\mathcal{M}}
\newcommand{\hiblop}{\mathcal{H}}
\newcommand{\fabricate}{\mathcal{G}}
\newcommand{\R}{\mathbb{R}}
\newcommand{\C}{\mathbb{C}}
\newcommand{\classscore}{\mathcal{S}}
\newcommand{\norm}[1]{\left\lVert#1\right\rVert}

\theoremstyle{definition}
\newtheorem{definition}{Definition}
\newtheorem{remark}{Remark}
\hypersetup{
  colorlinks=true,
  linkcolor=blue!60!black,
  citecolor=blue!60!black,
  urlcolor=blue!60!black
}
\newcommand{\safeincludegraphics}[2][]{%
  \IfFileExists{#2}{\includegraphics[#1]{#2}}{\fbox{\parbox[c][0.24\textheight][c]{0.85\linewidth}{\centering Missing figure: #2}}}%
}

\title{Photonic AI:\\ A Hybrid Diffractive--Holographic Neural System for\\
Passive Optical Real-Time Image Classification}
\author{
\textbf{Prakul Sunil Hiremath}\,\orcidlink{0009-0007-9744-3519} \\
\small Department of Computer Science and Engineering,\\
\small Visvesvaraya Technological University (VTU), Belagavi, India \\
\small Aliens on Earth (AoE) Autonomous Research Group, Belagavi, India \\
\small \href{mailto:prakulhiremath@vtu.ac.in}{\texttt{prakulhiremath@vtu.ac.in}} \\
\small \href{https://github.com/prakulhiremath}{\texttt{github.com/prakulhiremath}} \\
\small \href{https://aliensonearth.in}{\texttt{aliensonearth.in}}
}

\begin{document}
\date{}
\maketitle

\begin{abstract}
Edge intelligence is constrained by the energy and latency costs of shuttling data through electronic memory hierarchies. Optical systems offer a fundamentally different computational regime: once an input wavefront is launched into a structured medium, propagation, diffraction, and interference jointly enact a linear transformation whose cost is determined by wave physics rather than by clocked arithmetic. This paper develops a rigorous systems-level treatment of that regime and introduces a hybrid diffractive--holographic architecture for image classification. The proposed model couples a Diffractive Optical Neural Network (DONN) with a Holographic Interference-Based Learning (HIBL) operator---a formal map from digitally optimized phase distributions to physically realizable, fabrication-compatible interference patterns embeddable in passive optical elements. We express the full inference pipeline as a composition of encoding, phase modulation, free-space propagation, and intensity measurement operators, making explicit which quantities are learned, which are fixed by design, and where nonlinearity enters through photodetection. This operator-theoretic view resolves a persistent gap in the optical-ML literature between \emph{learning} a transformation and \emph{physically realizing} it. In physics-informed simulation on MNIST, a three-layer system with approximately 25,000 phase elements achieves 91.2\% test accuracy with propagation-limited nanosecond-scale latency. The primary contribution is not a performance claim but a precise computational framework: learned representations can be physically embedded into structured optical media so that inference is executed by wavefront transformation through a passive, fabricated object rather than by sequential electronic multiply--accumulate operations.
\end{abstract}

\section{Introduction}

\begin{figure}[t]
\centering
\safeincludegraphics[width=0.9\linewidth]{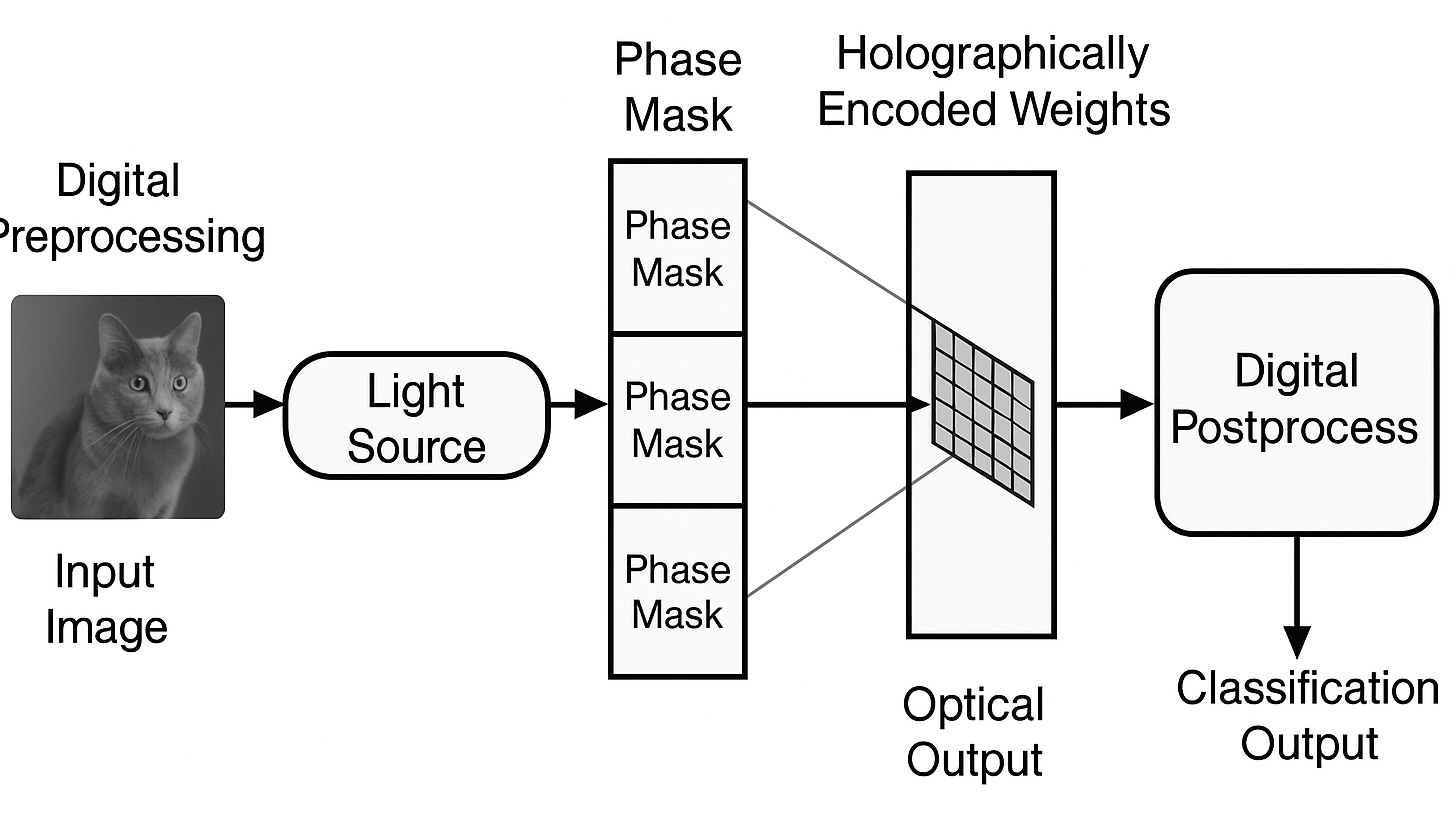}
\caption{Hybrid photonic neural system. Input data are encoded into a coherent optical field, transformed by cascaded diffractive layers optimized end-to-end, and realized as fabrication-compatible holographic phase structures via the HIBL operator. Class decisions are obtained from detector-plane optical energy integrated over disjoint output regions, with no floating-point arithmetic in the propagation path.}
\label{fig:architecture}
\end{figure}

The energy and latency budget of edge AI is dominated not only by arithmetic but by data movement. Modern inference engines repeatedly read weights from memory, stream activations across buses, and serialize computation through clocked pipelines. This von Neumann bottleneck is a structural feature of general-purpose digital hardware, not an incidental inefficiency amenable to incremental optimization. For compact, thermally constrained, or latency-critical platforms---autonomous microsystems, satellite payloads, implantable sensors---the mismatch between the computational workload and the available hardware resource is a design constraint, not merely a performance metric.

Optical computing is compelling in this context because it displaces part of the workload onto wave physics. A coherent field launched into a structured medium undergoes a transformation determined by geometry and material response, not by a fetched instruction stream. Within the constraints imposed by coherence, aperture, fabrication tolerance, and detector characteristics, this transformation is enacted in parallel, at the speed of light, and at a computational energy cost that scales with the illumination source rather than with the number of weights being applied. For inference workloads expressible as fixed linear operators followed by intensity measurement, this distinction is physically meaningful.

The realization of practical optical inference systems has proceeded along three partially disconnected trajectories. Diffractive deep neural networks (D2NNs) \citep{lin2018} established that cascaded passive phase masks can be optimized end-to-end to approximate target input--output mappings, demonstrating the representational capacity of free-space diffractive propagation for classification tasks. Integrated nanophotonic accelerators \citep{shen2017,miscuglio2020,feldmann2021} showed that optical interference can implement linear algebra---particularly matrix--vector multiplication---at chip scale with favorable energy-per-operation characteristics. Holographic and wavefront-shaping systems demonstrated that structured optical media can encode and reconstruct high-dimensional complex fields \citep{goodman}.

Despite these advances, a conceptual discontinuity persists across all three lines of work: the gap between the mathematical object that is optimized during training (a phase tensor or weight matrix) and the physical object that performs inference (a fabricated optical element). D2NN papers typically report accuracy for an abstract trained phase profile without formalizing the map from that profile to a manufacturable interference structure. Holographic systems are expert at encoding wavefronts into physical media but rarely connect that capability to an end-to-end trained inference pipeline. Photonic accelerators are functionally powerful but operate as execution engines rather than as passive, field-transforming structures.

This paper addresses the gap directly. We introduce a formal \emph{Holographic Interference-Based Learning} (HIBL) operator that maps a learned phase distribution to a physically recordable interference pattern consistent with holographic recording and fabrication constraints. HIBL is not a storage mechanism or a post-hoc encoding step; it is an operator in the computational model, and its inclusion changes the framing of optical inference from ``optimize a phase mask'' to ``learn a structured medium.''

The resulting system can be read as a two-stage compositional map:
\begin{equation}
\underbrace{x \;\mapsto\; U_L \;\mapsto\; \hat{y}}_{\text{inference pipeline}}
\qquad
\underbrace{\{\phi_\ell\} \;\mapsto\; \{H_{\phi_\ell}\} \;\mapsto\; \{T_{\phi_\ell}\}}_{\text{realization pipeline}}
\end{equation}
where the inference pipeline is the standard diffractive cascade, the realization pipeline is governed by HIBL and a fabrication map, and the two pipelines are coupled: what is optimized in training must be expressible in the realization pipeline.

The specific contributions of this work are:
\begin{itemize}[leftmargin=1.5em]
\item a physics-grounded operator decomposition of optical inference that separates learned, fixed, and emergent components of the computation;
\item the definition of HIBL as a formal operator from learned phase distributions to physically realizable interference patterns, closing the loop between digital training and passive optical embodiment;
\item a unified framing of the DONN+HIBL combination as a system that enacts \emph{inference as wavefront transformation through learned, fabricated structure}; and
\item a critically interpreted simulation study on MNIST that separates numerical performance from experimental feasibility and provides physically grounded interpretation of architectural design choices.
\end{itemize}

Section~\ref{sec:foundations} develops the optical and computational foundations. Section~\ref{sec:related} positions the work against the three paradigms identified above with explicit identification of the gap this work fills. Section~\ref{sec:method} formalizes the architecture. Sections~\ref{sec:experiments}--\ref{sec:results} present the experimental setup and results, and Section~\ref{sec:discussion} addresses limitations and implications.

\section{Optical Foundations of Computation}
\label{sec:foundations}

A rigorous treatment of optical inference requires more than analogy between optical propagation and matrix multiplication. It requires a precise account of (i) what class of transformations free-space optics can enact, (ii) where nonlinearity enters and how it interacts with learning, and (iii) why the physical instantiation of a transformation is different in kind from its digital execution. We develop each of these in turn.

\subsection{Scalar wave propagation as a linear operator}

A monochromatic scalar optical field at transverse plane $(x, y)$ is described by a complex amplitude $U(x, y) \in \C$, where the physical observable---intensity---is the squared modulus $I(x,y) = |U(x,y)|^2$. The complex representation is not merely a notational convenience; it encodes both amplitude and phase, and phase is the primary degree of freedom through which passive optical elements perform computation.

Free-space propagation from plane $z = 0$ to plane $z = d$ is, under the scalar paraxial approximation, a \emph{linear} operator on $U$. Its most computationally explicit form is the angular spectrum method (ASM):
\begin{equation}
U(x,y,d) = \InvFourier \left\{ \Fourier\left[U(x,y,0)\right] \cdot H(f_x,f_y,d) \right\},
\label{eq:asm}
\end{equation}
where $(f_x, f_y)$ are spatial frequencies conjugate to $(x,y)$, and
\begin{equation}
H(f_x,f_y,d) = \exp\!\left(j 2\pi d \sqrt{\lambda^{-2} - f_x^2 - f_y^2}\right)
\label{eq:transfer}
\end{equation}
is the free-space transfer function for wavelength $\lambda$, subject to the evanescent-wave cutoff $f_x^2 + f_y^2 \leq \lambda^{-2}$. Equation~\eqref{eq:asm} is important because it makes the computational structure explicit: propagation is a global linear operator \emph{diagonal in the spatial frequency domain}. Its action mixes all spatial positions of the input field through a phase-only spectral filter. Unlike a digital matrix--vector product, this mixing does not access stored weight values; it is enacted by geometry and the wave equation.

The Fresnel approximation, valid when propagation distance satisfies $d \gg (x^2 + y^2)_\mathrm{max}/\lambda$, simplifies \eqref{eq:transfer} to a quadratic phase factor and renders propagation as a scaled Fourier transform. For the propagation distances and spatial scales typical in free-space diffractive systems, the full ASM is preferable in simulation as it does not impose a paraxial constraint on spatial frequencies.

\begin{remark}
The ASM propagator $\asm_d$ is unitary on the space of propagating modes (those satisfying $f_x^2 + f_y^2 < \lambda^{-2}$) and strictly attenuates evanescent components. It therefore preserves the total power of propagating fields and introduces no gain, making it a faithful model for passive optical systems.
\end{remark}

\subsection{Phase modulation and its expressive role}

A thin phase-only optical element with profile $\phi(x,y) \in [0, 2\pi]$ acts on an incident field by pointwise multiplication:
\begin{equation}
U^{+}(x,y) = U^{-}(x,y) \cdot \exp\!\left(j\phi(x,y)\right).
\label{eq:phase_mod}
\end{equation}
This operation is \emph{local} in the spatial domain---each pixel of the mask acts independently on the field at that location---but becomes \emph{nonlocal} after subsequent propagation, since the ASM operator mixes spatial frequencies globally. The combination of local modulation and global mixing through alternating application of $\phasemap_\phi$ and $\asm_d$ constitutes the fundamental computational primitive of diffractive optical networks.

From a function approximation perspective, the expressive capacity of this primitive warrants careful analysis. A single phase layer followed by propagation and detection is equivalent, in the Fresnel regime, to computing a weighted coherent sum over the input field, followed by pointwise intensity measurement. The weights are determined by the phase profile and the propagation geometry and are not independently adjustable. Multiple such layers, however, can compose to produce transformations of substantially higher rank. Specifically, for $L$ layers with phase profiles $\phi_1, \ldots, \phi_L$ separated by propagation distances $d_1, \ldots, d_L$, the composite transformation maps the input complex field to an output intensity distribution through an operator of the form
\begin{equation}
\asm_{d_L} \circ \phasemap_{\phi_L} \circ \cdots \circ \asm_{d_1} \circ \phasemap_{\phi_1},
\label{eq:composite}
\end{equation}
followed by $|\cdot|^2$. The effective parameter space has dimension $\sum_\ell N_\ell$ where $N_\ell$ is the number of phase pixels in layer $\ell$, but the realized transformation is implicitly constrained by the structure of wave propagation: the set of achievable output distributions is a strict subset of all measurable functions on the detector plane.

Understanding this constraint is important for setting appropriate expectations about what optical inference can and cannot approximate. For sufficiently smooth targets and appropriate choice of architecture (layer count, propagation distances, spatial sampling), the achievable function class is rich enough to implement accurate classifiers for structured visual tasks \citep{lin2018}. Whether this richness extends to arbitrary targets remains an open theoretical question and is not addressed by the present work.

\subsection{Interference as a computational primitive}

Interference---the coherent superposition of field components---is not an incidental optical phenomenon but the mechanism by which diffractive systems produce class-selective output patterns. When two field components $U_1$ and $U_2$ overlap at the detector plane, the observed intensity is
\begin{equation}
|U_1 + U_2|^2 = |U_1|^2 + |U_2|^2 + 2\,\mathrm{Re}\{U_1 U_2^*\},
\label{eq:interference}
\end{equation}
where the cross term $2\,\mathrm{Re}\{U_1 U_2^*\}$ encodes the relative phase between components and can be either constructive or destructive depending on that phase difference. A well-designed diffractive system steers energy toward specific detector regions for a given input class by arranging interference patterns to be constructive at the correct output location and destructive elsewhere.

This makes the role of phase profiles precise: they are not analog weight values in the sense of a standard neural network, but rather shapers of the spatial frequency distribution that arrives at the detector plane. The ``weights'' of the optical computation are implicitly distributed across the interference structure of the entire propagating field. This has a direct consequence for expressiveness: a single complex-valued output at the detector integrates contributions from the entire input field through the global mixing of propagation. No spatial locality constraint limits which input regions can influence which output regions, which is different from convolutional processing and partially analogous to a fully connected layer.

\subsection{Nonlinearity, detection, and the hybrid inference model}

A fundamental point---often understated in optical-ML literature---is that passive coherent free-space propagation is \emph{linear in complex amplitude}. The field evolves according to the linear wave equation; no nonlinearity is introduced by propagation through a fixed phase medium. The principal nonlinearity of an optical inference pipeline arises at photodetection:
\begin{equation}
\measureop: U(x,y) \mapsto |U(x,y)|^2.
\label{eq:measurement}
\end{equation}
This measurement operation is quadratic in field amplitude, which is why intensity-domain readout can distinguish between fields that differ only in phase---a distinction inaccessible to a linear operation. Subsequent electronic processing (thresholding, argmax, softmax approximation) adds further nonlinearity, but at a computational scale orders of magnitude smaller than the linear transforms enacted by propagation.

This analysis identifies a structural truth about diffractive optical systems: they are not general-purpose approximators of arbitrary nonlinear functions over input intensity patterns. Rather, they are efficient realizers of \emph{structured linear transforms on complex fields}, followed by a fixed quadratic measurement. The task of learning is therefore to identify a phase structure $\{\phi_\ell\}$ such that the resulting measurement operator correctly separates class-conditional input distributions. Tasks for which useful class discrimination can be accomplished within this constrained functional form are natural targets for optical inference.

\begin{remark}
The absence of in-propagation nonlinearity has an important implication for depth. In electronic networks, depth is used to compose nonlinear functions and thereby expand the function class beyond what a single layer can approximate. In diffractive optical systems, depth instead increases the complexity of the realizable linear transform (by composing more alternating modulation--propagation stages), while nonlinearity is deferred to detection. Depth and detection are therefore not substitutes but serve distinct roles.
\end{remark}

\subsection{Physical computation versus von Neumann execution}

Electronic neural networks, whether running on CPUs, GPUs, or dedicated accelerators, are implemented on architectures that separate memory from compute. Weights are stored in memory arrays and streamed to arithmetic units at inference time; the computation is a sequence of multiply--accumulate operations clocked by an execution engine. The cost of inference is therefore dominated not merely by the number of operations but by the energy and latency of data movement---reading, transporting, and writing tensor elements across a memory hierarchy.

Passive diffractive optical systems are architecturally distinct. A trained model is not a collection of stored numbers applied to inputs by an execution engine; it is a geometrically structured object that \emph{is} the transformation. When an input wavefront interacts with a fabricated optical element, the transformation is enacted by the physical response of the medium to the incident field---by wave propagation, diffraction, and interference---rather than by fetched instructions. The execution engine is, in a precise sense, replaced by wave physics.

This distinction motivates the central framing of the present work: \emph{computation as propagation through learned, physically embedded structure}. The challenge is then not merely to optimize a phase distribution in simulation, but to define rigorously how that distribution is expressed as a physically realizable and stable optical object. This is precisely the role of the HIBL operator introduced in this paper.

\section{Related Work and Conceptual Positioning}
\label{sec:related}

Prior work on optical inference can be organized into three paradigms, each addressing a different aspect of the problem and each carrying a different limitation relevant to the goals of this paper.

\subsection{Diffractive optical neural networks}

The demonstration of all-optical machine learning using diffractive deep neural networks \citep{lin2018} established the viability of passive phase masks as computational layers. In that framework, each diffractive layer is parameterized by a phase profile, end-to-end gradient-based optimization is used to train these profiles for a target task, and inference is performed entirely optically. Experimental demonstrations on MNIST-class tasks confirmed that passive optical systems can achieve meaningful classification accuracy.

The intellectual importance of D2NNs is their identification of differentiable wave propagation as the bridge between optical physics and machine learning methodology. By treating the ASM as a differentiable operator, Lin et al.\ enabled backpropagation through the physics of free-space diffraction.

However, the D2NN literature has a structural limitation that is important to name precisely: the trained object is a \emph{phase tensor}, and the mapping from that tensor to a \emph{stable, fabrication-compatible physical element} is treated as an implementation detail rather than a component of the computational model. Fabrication of diffractive elements is described---via spatial light modulators, 3D printing, or lithography---but the operator that maps a continuously valued trained phase to a quantized, patterned, physically recordable structure is not formalized. This omission matters because that map has a non-trivial kernel (many distinct phase profiles may produce the same physical element) and a restricted range (not all phase profiles are physically realizable to arbitrary precision). Any performance claimed for a trained phase mask implicitly assumes this map is identity; in practice, it is not. The HIBL operator introduced in this paper explicitly addresses this gap.

\subsection{Photonic accelerators and integrated tensor cores}

A second paradigm uses integrated photonics to accelerate linear algebra. Programmable interferometric meshes of Mach--Zehnder interferometers \citep{shen2017} implement unitary matrix--vector products by controlling the phase and splitting ratios of integrated beamsplitters. Wavelength-division multiplexing architectures \citep{miscuglio2020} and phase-change-material tensor cores \citep{feldmann2021} extend this to parallel convolutional processing or dense multiply--accumulate operations with favorable energy-per-operation characteristics.

These systems are powerful, but they belong to a different computational regime from the passive free-space systems considered here. Photonic accelerators are \emph{execution-centric}: they are reconfigurable hardware engines that apply programmable weight values to input data, much as a digital processor applies stored weights but using optical interference rather than transistor switching. Weights are loaded into modulator settings; inference for different inputs uses the same active hardware.

By contrast, passive diffractive inference \emph{embeds} the transformation into the geometry of a static medium. There is no analog of weight loading; the ``weights'' are determined at fabrication time and persist in the physical structure of the element. The distinction is not merely energetic; it reflects a different model of what it means for a system to ``know'' a transformation. This work is not competitive with photonic accelerators for applications requiring flexibility or weight update; rather, it targets applications where the transformation is stable and where passive, low-overhead inference is preferred.

\subsection{Holographic and wavefront-shaping systems}

Holography provides the physical language for encoding arbitrary complex wavefronts into a recording medium through interference. Classical holographic memory \citep{yariv}, optical correlation filters, and more recent wavefront-shaping systems for scattering media all share the principle that a targeted optical response can be encoded in a physical structure by recording the interference pattern of an object beam and a reference beam. Reconstruction from the reference beam reproduces the object field.

The capability demonstrated by holographic systems---encoding and reliably reproducing high-dimensional complex fields---is directly relevant to the problem of realizing a trained diffractive network. A learned phase profile $\phi(x,y)$ can be viewed as specifying a desired optical response; holographic recording provides the physical mechanism by which that response is stably embedded in a fabricatable medium.

What holographic and wavefront-shaping literature does not provide, however, is an end-to-end connection from a \emph{learned} phase distribution to a \emph{trained inference system}. Holography is typically used to record a specific, known field; in optical inference, the field to be recorded is the \emph{output} of a gradient-based optimization over a classification objective, and the quality of the holographic encoding directly affects the achievable classification accuracy. Formalizing this connection---specifying the HIBL operator and characterizing how approximation errors in the holographic realization propagate into inference accuracy---is a central contribution of this paper.

\subsection{The gap this work fills}

The literature contains three strong but partially disconnected contributions: D2NNs show that passive optical layers can be trained; photonic accelerators show that optics can execute neural operations efficiently; and holographic systems show that complex optical fields can be stably embedded in physical media. What remains absent is a unified framework that (i) treats the map from trained phase distributions to physical realizations as an explicit operator in the computational model, (ii) uses this operator to close the loop between digital optimization and passive optical embodiment, and (iii) connects both the inference and the realization pipeline within a single formulation. This paper fills that gap.

\section{Methodology}
\label{sec:method}

\subsection{Input encoding and field representation}

Let $x \in \mathcal{X} \subset \R^{N_x \times N_y}$ denote an input image normalized to $[0,1]$. An optical encoder maps $x$ to an incident complex field
\begin{equation}
U_0(x,y) = A(x,y;\,x)\,\exp\!\left(j\theta_0(x,y)\right),
\label{eq:encoding}
\end{equation}
where $A$ is typically taken proportional to $\sqrt{x(x,y)}$ (square-root encoding, proportional to field amplitude) and $\theta_0$ is a fixed encoding phase, set to zero in the present study. The normalization $\norm{U_0}^2 = \text{const}$ ensures that all inputs carry the same total optical power, decoupling intensity-based class decisions from illumination strength.

We assume monochromatic, spatially coherent illumination, which ensures that the field evolves deterministically according to \eqref{eq:asm} and that interference effects are fully developed. This is an idealization: real sources have finite coherence length and bandwidth. The implications of this assumption are discussed in Section~\ref{sec:discussion}.

\subsection{Layer operators}

\begin{definition}[Phase modulation operator]
The $\ell$th diffractive layer is parameterized by a phase profile $\phi_\ell \in [0,2\pi]^{N_x \times N_y}$. Its action on an incident field $U$ is
\begin{equation}
\phasemap_{\phi_\ell}(U) = U \odot \exp(j\phi_\ell),
\label{eq:phiop}
\end{equation}
where $\odot$ denotes pointwise multiplication in the spatial domain.
\end{definition}

\begin{definition}[Propagation operator]
Free-space propagation over distance $\Delta z_\ell$ is the linear operator
\begin{equation}
\asm_{\Delta z_\ell}(U) = \InvFourier\!\left\{\Fourier[U]\cdot H(f_x,f_y,\Delta z_\ell)\right\},
\label{eq:pop}
\end{equation}
with transfer function $H$ as defined in \eqref{eq:transfer}.
\end{definition}

For an $L$-layer system, the complex field at the detector plane is
\begin{equation}
\boxed{U_L = \asm_{\Delta z_L} \circ \phasemap_{\phi_L} \circ \cdots \circ \asm_{\Delta z_1} \circ \phasemap_{\phi_1}(U_0).}
\label{eq:full_comp}
\end{equation}
This composition is the core computational model. The learned quantities are the phase profiles $\{\phi_\ell\}_{\ell=1}^{L}$; the design parameters $\{\Delta z_\ell\}$, $\lambda$, $N_x$, $N_y$, and the spatial sampling pitch are fixed before training.

\subsection{Measurement operator and implicit nonlinearity}

The passive optical stack in \eqref{eq:full_comp} is linear in complex field amplitude. The inference pipeline becomes nonlinear only through the measurement operator
\begin{equation}
\measureop(U) = |U|^2,
\label{eq:measurement2}
\end{equation}
followed by region-wise integration. Class scores are defined as
\begin{equation}
s_c = \classscore_c(U_L) = \int_{\Omega_c} \measureop(U_L)(x,y)\,dx\,dy,
\label{eq:score}
\end{equation}
where $\{\Omega_c\}_{c=1}^{C}$ are disjoint detector regions assigned to the $C$ classes. The predicted class is
\begin{equation}
\hat{y} = \arg\max_{c \in \{1,\ldots,C\}} s_c.
\label{eq:decision}
\end{equation}
This is the only decision-level computation required of the electronic subsystem; everything upstream of the photodetectors is realized optically. The system is therefore a genuinely hybrid pipeline: optical propagation executes the expensive high-dimensional linear mixing, while terminal electronics handle low-bandwidth detection and discrete decision logic.

\subsection{Loss and gradient flow}

Training minimizes a cross-entropy loss between the class scores $\{s_c\}$ and the one-hot ground-truth label $y$. Since $s_c$ is a differentiable function of $U_L$ and $U_L$ is a differentiable function of $\{\phi_\ell\}$ (via the ASM operators, implemented numerically with FFT), backpropagation through the optical stack is straightforward. The gradient of the loss with respect to $\phi_\ell$ is computed by differentiating through alternating FFT, spectral filter, and IFFT operations. This differentiability is what makes end-to-end optical network training feasible.

\subsection{HIBL as a realization operator}

\begin{definition}[HIBL operator]
Let $\phi(x,y)$ denote a learned phase profile produced by optical training. The holographic interference-based learning operator $\hiblop$ maps $\phi$ to a physically recordable interference pattern:
\begin{equation}
\hiblop: \phi(x,y) \;\mapsto\; H_\phi(x,y) = |O_\phi(x,y) + R(x,y)|^2,
\label{eq:hibl}
\end{equation}
where
\begin{equation}
O_\phi(x,y) = A_o(x,y)\,\exp\!\left(j\phi(x,y)\right)
\end{equation}
is an object beam encoding the learned phase, and
\begin{equation}
R(x,y) = A_r(x,y)\,\exp\!\left(j\theta_r(x,y)\right)
\end{equation}
is a known reference beam with fixed geometry.
\end{definition}

Expanding \eqref{eq:hibl}:
\begin{equation}
H_\phi(x,y) = A_o^2 + A_r^2 + 2 A_o A_r \cos\!\left(\phi(x,y) - \theta_r(x,y)\right).
\label{eq:hibl_expanded}
\end{equation}
The third term encodes the learned phase $\phi$ as an amplitude-modulated fringe pattern. The reference beam geometry $\theta_r$ and the amplitude profiles $A_o$, $A_r$ are design parameters of the recording setup. Equation~\eqref{eq:hibl_expanded} makes clear that $\hiblop$ is not a lossless embedding: the dc bias terms $A_o^2 + A_r^2$ and the amplitude modulation of the fringe reduce the fidelity with which $\phi$ is recovered when the hologram is illuminated by the reconstruction beam. Minimizing this fidelity loss is a materials and recording-design problem that we parameterize but do not fully solve in the present work.

The fabrication operator $\fabricate$ then maps the recorded interference pattern to a physically instantiated transmittance or phase response:
\begin{equation}
T_\phi = \fabricate(H_\phi).
\end{equation}
In practice, $\fabricate$ captures exposure--response nonlinearity, phase quantization, substrate index profile, and any calibration or bleaching steps required to convert intensity patterns to phase-modulating media. The pair $\fabricate \circ \hiblop$ defines the complete realization pipeline.

The two-stage structure of the full system is then:
\begin{equation}
\underbrace{x \;\xrightarrow{\text{encode}}\; U_0 \;\xrightarrow{(\ref{eq:full_comp})}\; U_L \;\xrightarrow{\measureop}\; \{s_c\} \;\xrightarrow{\arg\max}\; \hat{y}}_{\text{inference}}
\qquad
\underbrace{\phi_\ell \;\xrightarrow{\hiblop}\; H_{\phi_\ell} \;\xrightarrow{\fabricate}\; T_{\phi_\ell}}_{\text{realization, per layer}}
\end{equation}
with the constraint that inference is evaluated under $T_{\phi_\ell}$ rather than $\phi_\ell$ directly. In this paper, we evaluate inference with $\phi_\ell$ (ideal mask), which provides an upper bound on performance, while the realization pipeline is studied conceptually. Quantifying the accuracy degradation from $\phi_\ell \to T_{\phi_\ell}$ is a principal direction for future experimental work.

\subsection{Design parameters and assumptions}

The present formulation assumes:
\begin{enumerate}[leftmargin=1.5em,label=(\roman*)]
\item \emph{Monochromatic, spatially coherent illumination.} This ensures fully developed interference and deterministic propagation. Real sources with finite coherence partially degrade fringe visibility and thus the fidelity of the HIBL embedding.
\item \emph{Scalar diffraction.} Polarization effects are neglected, which is valid for paraxial angles and isotropic media.
\item \emph{Ideal phase-only modulation.} Amplitude modulation by the mask is assumed negligible. In practice, spatial light modulators or etched diffractive elements introduce coupling between phase and amplitude.
\item \emph{Perfect alignment.} Layer-to-layer axial spacing is assumed exact. Misalignment introduces phase errors that are equivalent to perturbations in $\phi_\ell$.
\item \emph{Noiseless detection.} Shot noise and read noise at the detector are not modeled. For high-brightness illumination, shot noise is negligible; for low-power or miniaturized systems, it may not be.
\end{enumerate}
These assumptions define the operating regime within which the reported accuracy is valid. They are not hidden approximations; they are the boundary conditions of the model, and departures from each condition can be analyzed as perturbations of the ideal system.

\section{Experimental Setup}
\label{sec:experiments}

The proposed framework is evaluated on MNIST as a controlled benchmark for optical classification. MNIST is chosen not because it is a demanding task but because it is sufficiently structured to isolate the behavior of the diffractive optical stack: the digit images exhibit spatial regularity and class-conditional structure that can be captured by a small number of diffractive layers, allowing clear attribution of performance to architectural choices rather than to dataset complexity.

The simulated system uses three diffractive phase layers, each with spatial resolution matching the input grid, yielding approximately 25,000 total optimizable phase values. Input digits are encoded into a coherent optical field via square-root amplitude encoding, propagated through the learned masks using the angular spectrum method, and measured on a detector plane partitioned into ten class regions. The predicted class is determined by maximum integrated optical energy.

Propagation distances and spatial sampling pitch are chosen to ensure that each layer introduces significant diffraction---that is, that the far-field mixing from each modulation stage is substantial relative to the input aperture. This is the regime in which the alternating modulation--propagation architecture has maximum expressive capacity; if propagation distances are too short, each layer acts nearly locally and the composition reduces to a single effective layer.

Training uses gradient descent with the Adam optimizer, differentiating through the ASM propagation via FFT-based automatic differentiation. The learning rate is decayed on a cosine schedule over 100 epochs with batch size 128.

Reported latency is optical propagation time through the three-layer stack and must be interpreted precisely: it is the time for a launched wavefront to traverse the optical path, which at the speed of light through centimeter-scale propagation distances corresponds to nanosecond-scale transit. This is not end-to-end system latency. Source modulation, mechanical alignment, photodetection, and I/O contribute additional latency at the system level.

\begin{figure}[t]
\centering
\safeincludegraphics[width=0.45\linewidth]{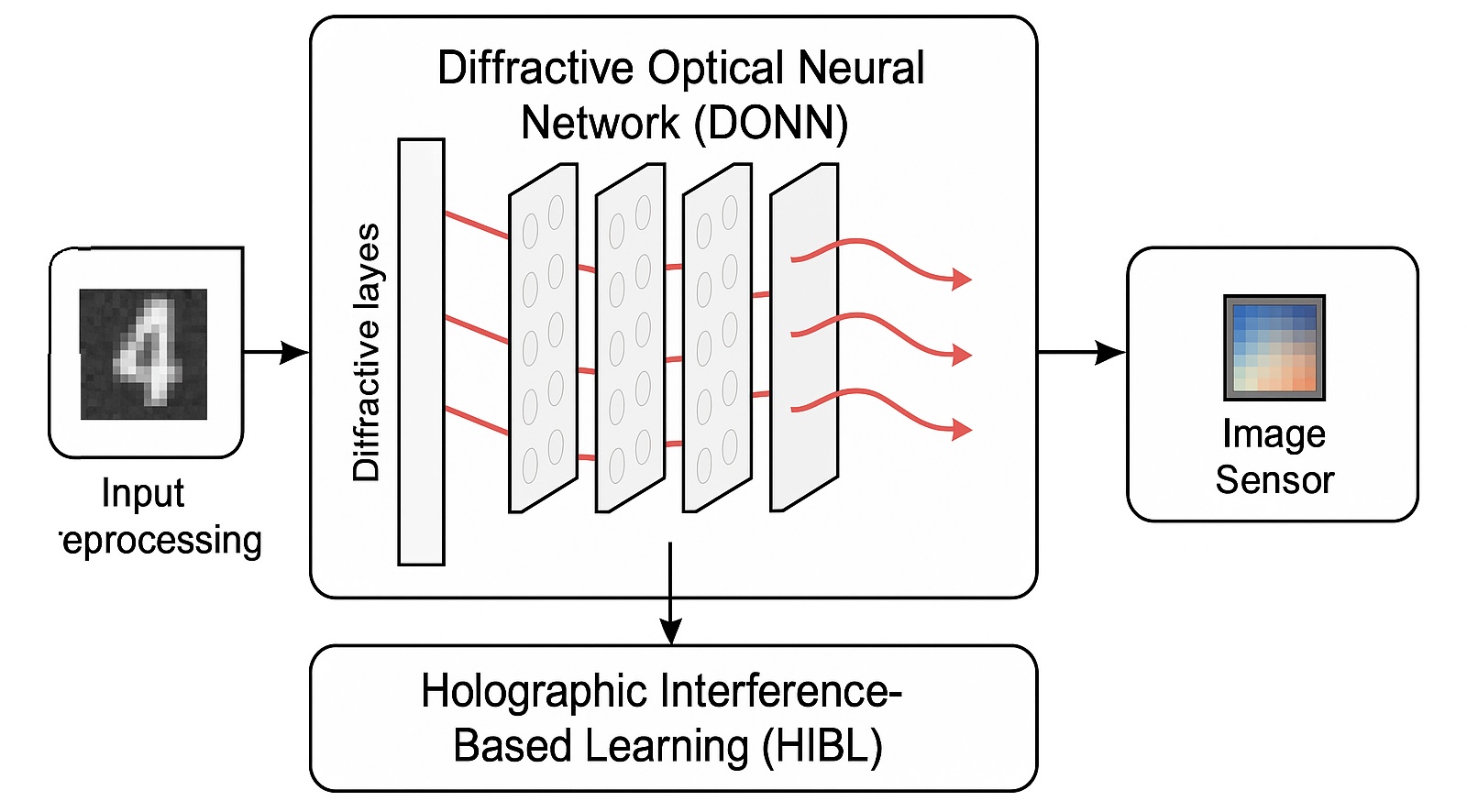}
\caption{Illustrative phase realization. \emph{Left:} Input digit intensity, encoded as the amplitude of a coherent field. \emph{Right:} Learned phase profile over $[0, 2\pi]$, interpreted through the HIBL operator $\hiblop$ as an interference fringe pattern embeddable in a physical recording medium. The fringe structure encodes the learned phase as a spatially varying modulation of the cosine term in \eqref{eq:hibl_expanded}.}
\label{fig:phasemask}
\end{figure}

Table~\ref{tab:performance} summarizes the principal simulation metrics.

\begin{table}[h]
\centering
\small
\begin{tabular}{p{0.34\linewidth}p{0.14\linewidth}p{0.36\linewidth}}
\toprule
Metric & Value & Interpretation \\
\midrule
Test accuracy & 91.2\% & MNIST test set, physics-informed simulation \\
Propagation latency & $\sim$ns & Optical transit time through the three-layer stack \\
Computational energy (propagation) & Negligible & No electronic MAC operations in the optical path \\
Phase elements & $\approx 25\mathrm{k}$ & Three diffractive layers, jointly optimized \\
\bottomrule
\end{tabular}
\caption{Performance summary for the simulated hybrid diffractive--holographic classifier. ``Negligible'' refers specifically to computational energy expended during passive propagation; total system energy including source, detection, and control electronics is non-negligible and task-dependent.}
\label{tab:performance}
\end{table}

\section{Results and Interpretation}
\label{sec:results}

\begin{table}[h]
\centering
\small
\begin{tabular}{p{0.26\linewidth}p{0.2\linewidth}p{0.2\linewidth}p{0.2\linewidth}}
\toprule
System & Medium & Latency Regime & Operational Character \\
\midrule
Hybrid photonic AI (this work) & Structured optical + electronic readout & ns-scale propagation & Passive fixed transform \\
MobileNetV1 & Electronic & ms-scale & Programmable digital inference \\
Jetson Nano deployment & Electronic & ms-scale & Embedded GPU execution \\
\bottomrule
\end{tabular}
\caption{Qualitative comparison with representative electronic deployment regimes. The contrast is architectural rather than task-normalized. The purpose is to distinguish execution modality---repeatedly applied digital weights versus passively enacted optical transformation---not to assert universal superiority of the optical approach.}
\label{tab:comparison}
\end{table}

\begin{figure}[t]
\centering
\safeincludegraphics[width=0.7\linewidth]{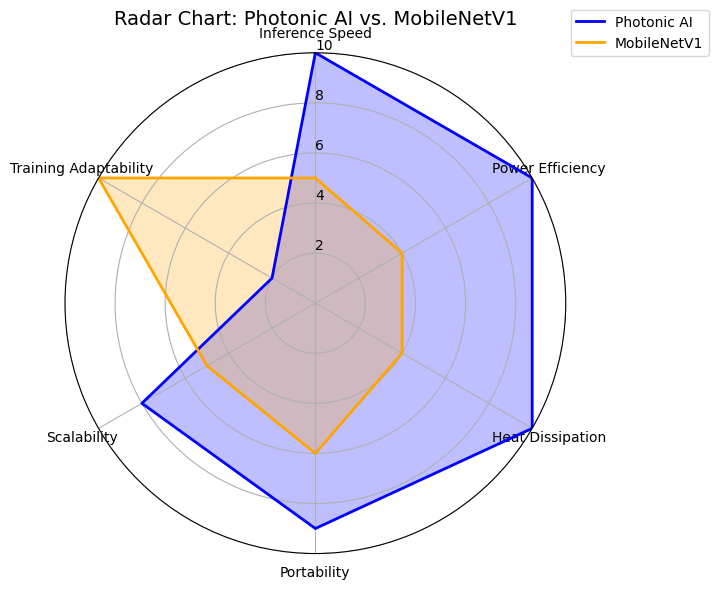}
\caption{Deployment-level comparison between passive optical inference and electronic models. The salient distinction is not latency alone, but where the transformation is instantiated: in a geometrically structured optical medium versus in repeatedly executed digital arithmetic. This difference has consequences for power, latency, reconfigurability, and the physical permanence of the model.}
\label{fig:radar}
\end{figure}

\subsection{Classification accuracy and what it implies}

The simulated three-layer system achieves 91.2\% test accuracy on MNIST. This result has a specific interpretation within the framework developed here: it demonstrates that the constrained functional class accessible to passive diffractive propagation---structured linear transforms on complex amplitude followed by intensity measurement---is sufficient to separate the class-conditional distributions of handwritten digit images. MNIST is sufficiently low-complexity that this sufficiency is perhaps not surprising, but it is also non-trivial: the optical system has access only to spatial interference patterns shaped by the learned phase profiles, with no intermediate nonlinear activation functions.

The gap between 91.2\% and the near-100\% accuracy achievable by deep electronic networks on MNIST should be understood as a consequence of the functional constraint, not of insufficient optimization. The optical system operates in a genuinely restricted hypothesis class. Closing that gap would require either more layers (and thus more composable mixing), wider apertures (richer spatial frequency content), or richer modulation (amplitude as well as phase).

\subsection{Why three layers? The physics of diminishing returns with depth}

The three-layer configuration is a meaningful operating point, and the reasoning is grounded in the physics of alternating modulation--propagation architectures. Each additional layer contributes another cycle of local phase shaping followed by global frequency mixing. The first few layers produce the largest increment in expressive capacity because they convert a highly correlated input field (the encoded digit) into a spatially redistributed intermediate field with more broadly spread spectral content. Subsequent layers refine this distribution but operate on a field that is already partially disordered; the marginal information gain from another mixing stage decreases.

On MNIST specifically, the first layer already redistributes most of the class-dependent spatial energy content of the digit images, and layers two and three perform increasingly fine-grained refinement of the class-separating output pattern. The detector geometry---ten fixed spatial regions---introduces an output bottleneck that further limits the benefit of additional layers. Once the output intensity distribution is spatially organized enough to produce clear class-region maxima, additional depth primarily reduces cross-class leakage rather than establishing new discriminative structure.

This analysis has a fabrication implication: systems designed for tasks of moderate complexity can be built with few layers, reducing alignment sensitivity, substrate count, and fabrication cost. More complex visual tasks would require a principled capacity analysis to determine whether additional layers or larger apertures provide better returns.

\subsection{Robustness and its physical meaning}

Consistency of performance across training runs in simulation reflects a property of the optimization landscape: the learned phase profiles do not rely on extremely delicate interference cancellations at every spatial point in the field. Rather, the learned structures direct energy toward class-specific detector regions through broadly cooperative interference patterns that are stable under small perturbations to the phase values.

It is important to distinguish numerical robustness from experimental robustness. In simulation, a phase perturbation of $\delta\phi$ in layer $\ell$ induces a field perturbation at the detector proportional to $\delta\phi$ times the sensitivity of the propagated field to that layer's profile. In an experiment, the same perturbation arises from fabrication error, alignment drift, or thermal expansion. Numerically robust solutions are likely better starting points for fabrication---their performance degrades more gracefully under perturbation---but experimental validation of this claim requires physical prototyping beyond the scope of this paper.

\subsection{Energy interpretation}

The system consumes no electronic energy in the optical propagation path, since passive propagation through a fabricated element requires only input illumination. The relevant comparison to electronic inference is therefore in terms of \emph{computational energy}---the energy expended per inference operation---rather than total system power. The illumination source, photodetectors, and control electronics contribute a fixed overhead that is largely independent of input complexity, unlike digital inference engines whose power scales with model size and arithmetic throughput. For applications with high-throughput inference demands (many inferences per second) or low-duty-cycle sensing (intermittent illumination), the energy profile of passive optical inference may differ substantially from its electronic counterparts in operationally important ways. We refrain from quantitative energy comparisons here, as they require explicit system-level specifications beyond the scope of this simulation study.

\section{Discussion}
\label{sec:discussion}

\subsection{Physical embedding as a model of inference}

The most useful interpretation of this work is as a statement about where a learned model can reside. In standard machine learning hardware, the model resides in a memory array and is instantiated by an execution engine at inference time---it is an informational object, repeatedly interpreted by a computational process. In the proposed framework, the model is partially resident in the geometry of a passive optical medium. Inference is a physical interaction between an input wavefront and a learned structure. The model does not need to be fetched, decoded, or scheduled; it is simply there, and the field encounters it.

This framing has an important implication for the concept of model ``deployment.'' A diffractive optical model is deployed by fabrication. Updates require re-fabrication (or, in partially reconfigurable systems, re-programming of active elements). The trade-off is therefore between fixity and efficiency: embedding a model in a passive medium achieves extreme inference speed and low computational overhead but sacrifices the on-the-fly updatability of digital systems. This is not a deficiency of the optical approach; it is the correct engineering trade-off for applications where the inference task is stable and the premium is on throughput, latency, or energy.

\subsection{Limitations and scope}

Several limitations define the scope of the current contribution:

\textbf{Fabrication sensitivity.} Small deviations in phase-mask thickness, refractive index, or surface figure produce phase errors at each pixel. For spatial light modulator implementations, pixel-to-pixel uniformity and addressing precision limit achievable phase fidelity. For lithographically patterned elements, etch depth uniformity is the dominant error source. In all cases, the aggregate effect of fabrication error can be modeled as a stochastic perturbation of $\phi_\ell$, and robustness-inducing regularization during training (e.g., adding phase noise during forward passes) is a straightforward mitigation.

\textbf{Coherence requirements.} Fully developed interference requires spatial coherence across the aperture and temporal coherence over path-length differences. Laser sources satisfy these requirements readily; LED or broadband sources require spatial filtering and narrow-band selection at the cost of power efficiency.

\textbf{Scaling to complex tasks.} Higher-resolution inputs and richer datasets require either larger apertures, finer spatial sampling, more layers, or more complex modulation. Each choice has fabrication and alignment implications. Extending the operator-theoretic framework to amplitude-and-phase modulation (complex spatial light modulators) or to wavelength-multiplexed multi-channel architectures is a natural generalization not addressed here.

\textbf{HIBL fidelity.} The HIBL operator as defined maps a continuous learned phase to an interference pattern that encodes that phase as a fringe modulation (see \eqref{eq:hibl_expanded}). The dc bias and the amplitude envelope of the fringe reduce the fidelity of phase reconstruction. Phase-conjugate readout or carrier-frequency holography can improve this, but detailed optimization of the recording geometry is left to future experimental work.

\textbf{Hybrid system overheads.} Source modulation, photodetection, and any electronic post-processing contribute non-negligible system-level energy and latency. Claims about the advantages of optical inference must be evaluated at the system level, not only at the propagation level.

\subsection{Toward a research program}

The framework introduced here suggests a concrete research program. Immediate experimental priorities include: (i) fabricating phase masks from trained profiles and measuring accuracy degradation relative to simulation, which directly quantifies the gap between ideal and realized HIBL; (ii) characterizing the sensitivity of accuracy to controlled phase perturbations, validating the robustness analysis of Section~\ref{sec:results}; and (iii) testing whether regularization during training (phase-noise injection, phase quantization) improves experimental accuracy as predicted.

Theoretical extensions include: a capacity analysis of the diffractive functional class as a function of layer count, propagation distance, and aperture; a formal characterization of the HIBL realization error and its propagation into inference accuracy; and the extension of the framework to partially reconfigurable systems (e.g., liquid-crystal-on-silicon backplanes) where inference-time weight updates are possible.

\section{Conclusion}

We have presented a hybrid diffractive--holographic neural system and, more importantly, a formal framework for interpreting what such a system computes. The central contribution is the HIBL operator, which maps learned phase distributions to physically realizable interference patterns and thereby closes the loop between digital optimization and passive optical embodiment. This addresses a conceptual gap in the optical-ML literature between learning an optical transformation and instantiating it as a stable physical object.

Within a physics-informed simulation on MNIST, the three-layer system achieves 91.2\% accuracy with propagation-limited nanosecond-scale latency and no computational energy expenditure in the optical path. The significance lies less in the benchmark accuracy itself than in the computational framing it supports: some learned models need not be executed as digital programs but can instead be materially embedded in structured optical substrates and evaluated by the physics of wavefront propagation through those structures. The challenge of realizing this framing in physical hardware---quantifying HIBL fidelity, characterizing fabrication sensitivity, and demonstrating accuracy parity between simulation and experiment---defines the research agenda to which this work contributes the necessary conceptual and operator-theoretic foundation.

\bibliographystyle{plainnat}

\end{document}